\documentclass[twocolumn,preprintnumbers,amsmath,amssymb,superscriptaddress]{revtex4}

\usepackage{graphicx}
\usepackage{dcolumn}
\usepackage{bm}
\usepackage{soul}
\usepackage{color}
\usepackage{epstopdf}
\usepackage[version=3]{mhchem}
\usepackage{lipsum}
\usepackage[outercaption]{sidecap}
\usepackage{floatrow}
\begin{document}

\title{Coupling between structural relaxation and diffusion in glass-forming liquids under pressure variation}
\author{Anh D. Phan}
\affiliation{Faculty of Materials Science and Engineering, Phenikaa Institute for Advanced Study, Phenikaa University, Hanoi 12116, Vietnam}
\email{anh.phanduc@phenikaa-uni.edu.vn}
\affiliation{Faculty of Computer Science, Artificial Intelligence Laboratory, Phenikaa University, Hanoi 12116, Vietnam}
\affiliation{Department of Nanotechnology for Sustainable Energy, School of Science and Technology, Kwansei Gakuin University, Sanda, Hyogo 669-1337, Japan}
\author{Kajetan Koperwas}
\affiliation{University of Silesia in Katowice, Institute of Physics, 75 Pulku Piechoty 1, 41-500, Chorzow, Poland}
\affiliation{Silesian Center
for Education and Interdisciplinary Research SMCEBI, 75 Pulku Piechoty 1a, 41-500, Chorzow, Poland}
\email{kajetan.koperwas@us.edu.pl}
\author{Marian Paluch}
\affiliation{University of Silesia in Katowice, Institute of Physics, 75 Pulku Piechoty 1, 41-500, Chorzow, Poland}
\affiliation{Silesian Center
for Education and Interdisciplinary Research SMCEBI, 75 Pulku Piechoty 1a, 41-500, Chorzow, Poland}
\author{Katsunori Wakabayashi}
\affiliation{Department of Nanotechnology for Sustainable Energy, School of Science and Technology, Kwansei Gakuin University, Sanda, Hyogo 669-1337, Japan}
\date{\today}

\date{\today}

\begin{abstract}
We theoretically investigate structural relaxation and activated diffusion of glass-forming liquids at different pressures using both the Elastically Collective Nonlinear Langevin Equation (ECNLE) theory and molecular dynamics (MD) simulation. An external pressure restricts local motions of a single molecule within its cage and triggers slowing down of cooperative mobility. While the ECNLE theory and simulation generally predict a monotonic increase of the glass transition temperature and dynamic fragility with pressure, simulation indicates a decrease of fragility as pressure above 1000 bar. The structural relaxation time is found to be linearly coupled with the inverse diffusion constant. Remarkably, this coupling is independent of compression. Theoretical calculations agree quantitatively well with simulations and are also consistent with prior works.
\end{abstract}

\maketitle

\section{Introduction}
Recently, molecular dynamics of glass-forming liquids during vitrification has been intensively investigated since physical mechanisms remain poorly understood despite a wide range of applications \cite{49,50,51,52,53}. When molten materials are cooled with a rapid rate to the temperatures below the melting temperature, their structure remains disordered, and they fall out of equilibrium. Molecular mobility in amorphous materials is characterized by diffusive transport of molecules and relaxation processes including structural/alpha and secondary relaxation. These relaxations are very sensitive to temperature and external pressures \cite{63,64,65,66}, which are two decisive factors in manufacturing and storage, as well as they are crucial for improving the solubility of products \cite{49,50}. Investigating the pressure and temperature dependence of (alpha) relaxation time and diffusion constant for glass formers is essential for industrial applications and acquires new knowledge of fundamental science.

Apart from experiments, there are several main approaches to investigate the molecular mobility in glass-forming liquids. Molecular dynamics (MD) simulations can capture effects of intermolecular interactions and penetration, geometrical structures, and rotational motions of molecules on the glass transition. However, the largest disadvantage of the simulations is their limited timescale of relaxation process (no longer than $10^6$ ps), which is much less than the experimental observation timescale (100 s). Meanwhile, the temperature dependence of the structural and secondary relaxation time at different pressures can be theoretically determined using the Elastically Collective Nonlinear Langevin Equation (ECNLE) theory \cite{61,62,2,6,7,10,8,42,11}. Since the original form of ECNLE is valid solely at ambient pressure, Phan and his coworkers \cite{42,11,61,62} proposed the method to extend its applicability for higher pressure, which is a density-to-temperature conversion (a thermal mapping) based on the thermal expansion process. The timescale predicted by the ECNLE theory ranges from from 1 ps to more than 100 s. In the ECNLE theory, an amorphous material is modeled as a fluid of spherical and impenetrable molecular particles, which means that the geometric factors and biological complexities are completely ignored. Nevertheless, this approach has showed a quantitative good agreement with various experiments for single- and multi-component systems \cite{42,11,61,62,38} including metallic glasses (a very recent work \cite{38}). In Ref. \cite{38}, we used the ECNLE theory to qualitatively elucidate nature of the reversible relaxation, compression-induced rejuvenation, and roles of precompression on the strain hardening of metallic glasses. Although oxide glasses have not been studied yet, one can expect this theoretical approach generally well with both inorganic and organic glasses.

In this paper, we combine MD simulations and the ECNLE theory to investigate the pressure and temperature dependences of structural relaxation time and diffusion constant. Based on these data, we determine dynamic fragility at different pressures. Simulation is quantitatively consistent with ECNLE calculations. The good accordance allows us to reveal a correlation between the relaxation and diffusion process. 

\section{Computational method}
We employ a recently proposed model of quasi-real liquid, i.e., the rhombus like molecules system (RLMS) \cite{9}. The advantage of RLMS is that, in contrast to standard simple-liquids systems, it exhibits a crucial feature of real material, which is structural anisotropy, keeping the simplicity of molecular architecture as much as possible. To achieve above, the  model molecules are composed of four identical atoms (of carbon mass), which are arranged in the shape of the rhombus. The bond lengths are identical and equal to 0.14982 nm (0.14 nm is a bond length for carbon atoms in a benzene ring). Additionally, to ensure the possibility of the creation of permanent dipole moments of different orientations, the angles between bonds are set to make one diagonal two times longer than the other. Consequently, the intermolecular interactions result from mutual interplay between 8 atoms of two different molecules. The interactions between non-bonded atoms as well as bonded atoms are set using parameters of the optimized potentials for liquid simulations (OPLS) for carbon atoms from aromatic ring \cite{12}. In this context, in order to ensure the maximal level of simplicity of RLMS, we decided not to add hydrogen atoms. This implies the redefinition for all atoms charges, which was set to 0.0$e$ ($e$ is an elementary charge). 

The functional form of the OPLS all-atom force field is mostly the same as that of AMBER force field \cite{39}, but parameters are determined by different optimizations to fit \emph{ab initio} molecular orbital calculations and experiments. The potential function in the OPLS all-atom force field includes van der Waals (described by the Lennard-Jones potential), electrostatic (given by the Coulomb potential), dihedral (represented by the Ryckaert-Bellemans potential), bond and bond angle terms. Since our atoms are neutral, there is no Coulomb interaction in our system.

The MD simulations for supercooled liquid at ambient pressure require the determination of the melting temperature. However, we use our previous results, where the melting conditions at 1000 bar have been determined \cite{13}. Subsequently, consistently with Ref. \cite{13}, we performed the cooling of liquid at constant pressure (NPT conditions) provided by the Nose-Hoover thermostat \cite{16,17,18} and Martyna-Tuckerman-Tobias-Klein barostat \cite{19,20}, which are implemented in the GROMACS software \cite{21,22,23}. Each simulation run lasts for a relatively long time, i.e., 10 ns, which means one billion of the time-steps, $dt=0.001$ ps. 

The first half of the simulation was devoted to equilibration of the system, whereas the volumetric data have been collected for the last 5 ns. However, at this point, a typical method for estimating the relaxation times from simulations of molecular dynamics, which is the analysis of the incoherent intermediate scattering function, requires simulations at constant temperature and volume (NVT). Therefore, we had used the determined volume under the constant pressure (NPT) condition, and we simulated RLMS at the respective NVT conditions. Then, the relaxation times, $\tau_\alpha$, are estimated on the base of incoherent intermediate scattering function of molecules centers of mass, whereas the diffusion constants, $D$, are obtained from mean-square displacement using GROMACS software. 

\section{ECNLE theory for molecular dynamics under compression}
In order to obtain more theoretical insights into calculations of the simulated glass-forming liquid at different pressures, we use ECNLE theory. This approach theoretically considers real amorphous materials by a hard-sphere fluid model \cite{2,3,4,7,8,10,6,42,11,61,62}. Two key parameters of the fluid are a particle diameter, $d$, and the number of particles per volume, $\rho$. Based on the Persus-Yevick (PY) theory \cite{1}, we obtain the static structure factor, $S(q)$, with $q$ being the wavevector, the direct correlation function, $C(q)=\left[S(q)-1 \right]/\rho S(q)$, and the radial distribution function, $g(r)$. The mobility of a tagged particle is mainly driven by nearest-neighbor interactions and an external pressure. According to a recent developed ECNLE theory \cite{61}, the free dynamic energy of the tagged particle under the pressure $P$ at temperature $T$ is

\begin{eqnarray}
\frac{F_{dyn}(r)}{k_BT} &=& \int_0^{\infty} dq\frac{ q^2d^3 \left[S(q)-1\right]^2}{12\pi\Phi\left[1+S(q)\right]}\exp\left[-\frac{q^2r^2(S(q)+1)}{6S(q)}\right]
\nonumber\\ &-&3\ln\frac{r}{d} + \frac{P}{k_BT/d^3}\frac{r}{d},
\label{eq:2}
\end{eqnarray}
where $k_B$ is Boltzmann constant, $r$ is the displacement of the particle, and $\Phi = \rho\pi d^3/6$ is the volume fraction. Recall that Eq. (\ref{eq:2}) is constructed by considering only translational motions. The first term corresponds to the caging constraint caused by nearest neighbors. The second term favors the delocalized state of particle or the ideal fluid state, while the last term is responsible for pressure effects. 

By using this free energy profile, we can obtain physical quantities of local cage-scale dynamics. When the density is sufficiently large ($\Phi \geq 0.432$) \cite{2,3,4,7,8,10,6,42,11,61,62}, an onset of particle localization occurs and the tagged particle is dynamically arrested within a cage formed by its neighbors. The local dynamics is characterized by an emergence of a barrier in $F_{dyn}(r)$. It is possible to consider a position of the first minimum of $g(r)$ as the particle cage radius, $r_{cage}$. The local minimum and maximum of the dynamic free energy are the localization length, $r_L$, and the barrier position, $r_B$, respectively. Then, we can calculate a jump distance $\Delta r =r_B-r_L$ and a local barrier height $F_B=F_{dyn}(r_B)-F_{dyn}(r_L)$.

When a particle wants to move over the particle cage, surrounding particles are required to rearrange for diffusing. The rearrangement of particles in the first shell leads to the surface expansion of the particle cage and it radially spreads out the remaining space. Suppose that the expansion is really small, one can approximately quantify the propagation by a displacement field $u(r)$. In bulk systems, an analytical form of the distortion field is calculated by Lifshitz's continuum mechanics analysis \cite{5}, which is
\begin{eqnarray} 
u(r)=\frac{\Delta r_{eff}r_{cage}^2}{r^2}, \quad {r\geq r_{cage}},
\label{eq:3}
\end{eqnarray}
where $\Delta r_{eff}$ is the amplitude of the cage expansion \cite{6,7}. This is
\begin{eqnarray} 
\Delta r_{eff} = \frac{3}{r_{cage}^3}\left[\frac{r_{cage}^2\Delta r^2}{32} - \frac{r_{cage}\Delta r^3}{192} + \frac{\Delta r^4}{3072} \right].
\end{eqnarray}
Since $u(r) \ll r_{cage}$, collective particles beyond the particle cage can be viewed as harmonic oscillators with a spring constant at $K_0 = \left|\partial^2 F_{dyn}(r)/\partial r^2\right|_{r=r_L}$ $\approx3k_BT/r_L^2$ and are harmonically vibrated with the amplitude $u(r)$. These calculations of the displacement field are similar to the phenomenological "shoving model" proposed by Dyre \cite{36,37}. Although Dyre derived the displacement field in Eq. (\ref{eq:3}), the amplitude of his $u(r)$ is an empirically adjustable parameter.

The total elastic energy of these oscillators, which is called the elastic barrier, is calculated to determine collective motion effects in relaxation processes. Based on the molecular Einstein perspective, we can calculate this elastic barrier as \cite{2,7,8,10,6,42,11,61,62}
\begin{eqnarray} 
F_{e} = 4\pi\rho\int_{r_{cage}}^{\infty}dr r^2 g(r)K_0\frac{u^2(r)}{2}. 
\label{eq:5}
\end{eqnarray}

To zeroth-order approximation, one can approximate $g(r)\approx 1$ and $\Delta r_{eff} \approx 3\Delta r^2/32r_{cage}$ \cite{2,7,8,10,6,42,11,61,62}. In addition, in prior works \cite{2,6,7}, an analytic formula for the dynamic shear modulus $G$ is derived as
\begin{eqnarray} 
G = \frac{9\Phi k_BT}{5\pi d r_L^2}. 
\label{eq:5-2}
\end{eqnarray}
Thus, these give
\begin{eqnarray} 
F_{e} = 12\Phi K_0\Delta r_{eff}^2\left(\frac{r_{cage}}{d} \right)^3 = 20\pi\frac{\Delta r_{eff}^2r_{cage}^3}{d^2} G.
\label{eq:5-1}
\end{eqnarray}

From this, one can clearly see that local and elastic barrier are physically distinct but strongly inter-correlated. However, the coupling between these two barriers is found to be non-universal due to chemical, biological, and conformational (shape) complexities \cite{11,8,61}. To address this non-universality, we suppose that the complexities modify a jump distance as $\Delta r \rightarrow \lambda\Delta r$, where we adjust a constant $\lambda$ to simultaneously provide the best theoretical description with simulations and experiments. This assumption implies the elastic barrier is scaled up to be $F_e \rightarrow \lambda^4 F_e$ \cite{11,8,61} and it can be used to determine the relative role of the collective motions in glassy dynamics. 

According to Kramer's theory, the structural relaxation time is
\begin{eqnarray}
\frac{\tau_\alpha}{\tau_s} = 1+ \frac{2\pi}{\sqrt{K_0K_B}}\frac{k_BT}{d^2}\exp\left(\frac{F_B+\lambda^4F_e}{k_BT} \right),
\label{eq:6}
\end{eqnarray}
where $K_B$=$\left|\partial^2 F_{dyn}(r)/\partial r^2\right|_{r=r_B}$ is absolute curvatures at the barrier position and $\tau_s$ is a short time scale of relaxation. The expression of $\tau_s$ is \cite{6,7,11,61,42}
\begin{eqnarray}
\tau_s=g^2(d)\tau_E\left[1+\frac{1}{36\pi\Phi}\int_0^{\infty}dq\frac{q^2(S(q)-1)^2}{S(q)+b(q)} \right],
\label{eq:8}
\end{eqnarray}
where $b(q)=1/\left[1-j_0(q)+2j_2(q)\right]$, $j_n(x)$ is the spherical Bessel function of order $n$, and $\tau_E \approx 0.5\times 10^{-13}$ s is the Enskog time scale \cite{2,6,7,42,11}.

Near the experimental $T_g$ (defined by $\tau_\alpha(T_g)=100$ s), ($F_e > F_B$) \cite{2,6}. Thus, we can crudely deduce Eqs. (\ref{eq:5-1}) and (\ref{eq:6}) to $\tau_\alpha \sim e^{GV_c/k_BT}$, here $V_c$ being a characteristic volume. This is consistent with the shoving model \cite{36}.

To compare our numerical calculations with experiments and simulations at ambient pressure, we use a density-to-temperature conversion (thermal mapping) \cite{42,11,61,62}, which is proposed using the thermal expansion process. The thermal mapping is expressed by \cite{42,11,61,62}
\begin{eqnarray}
T \approx T_0 - \frac{\Phi - \Phi_0}{\beta\Phi_0}.
\label{eq:7}
\end{eqnarray} 
where $\Phi_0 \approx 0.5$ is the characteristic volume fraction and $\beta \approx 12\times 10^{-4}$ $K^{-1}$ is a common value of the coefficient of linear expansion of many materials. We adjust $T_0$ and $a_c$ to obtain the best quantitative agreement between theory and experiment for $\tau_\alpha(T)$. Then, the hopping diffusion constant is calculated by
\begin{eqnarray}
D = \frac{(\lambda\Delta r)^2}{6\tau_\alpha}\equiv\frac{(\lambda\Delta r)^2}{6\tau_\alpha(T,P)}.
\label{eq:1}
\end{eqnarray}

It is important to note that the ECNLE calculations ignore effects of rotational motions. To be consistent these effects are also ignored in our calculations of $\tau_\alpha$ from simulations where we only consider the motion of the center of molecular mass.

\section{Results of diffusion and structural Relaxation}
To obtain the best fit between theory and simulations at ambient environment, key parameters for ECNLE calculations are $T_0 = 24$ $K$ and $\lambda \approx 3.39$ and we assume that these parameters are independent of pressure. It means the correlation between local and collective dynamics remains unchanged during compression. This assumption was used in Ref. \cite{61} and successfully described $\tau_\alpha(T,P)$ of curcumin, glibenclamide, and indomethacin when compared to experiments.

\begin{figure}[htp]
\center
\includegraphics[width=9cm]{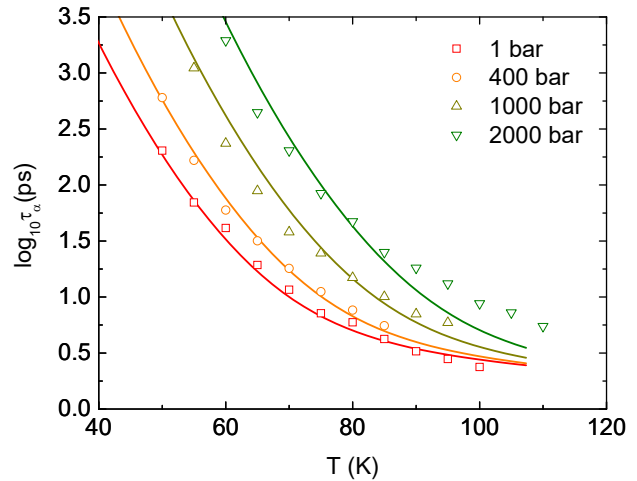}
\caption{\label{fig:1}(Color online) Logarithm of alpha relaxation time as a function of temperature at different pressures. Open points are simulations and solid curves correspond to ECNLE calculations.}
\end{figure}

Figure \ref{fig:1} show the temperature dependence of $\tau_\alpha$ under different compression conditions calculated using MD simulations and ECNLE theory. We find that $P/(k_BT/d^3)$ = 0, 0.4, 1.0, and 1.5 in the ECNLE calculations correspond to 1, 400, 1000, and 2000 bar, respectively. These findings indicate that one can approximate the particle diameter by $k_BT/d^3 \approx 1000$ bar for all pressures. According to this approximation, the atmospheric pressure is considered as $10^{-3}k_BT/d^3$, which is close to 0. It is quite plausible since there is no difference between theoretical calculations using $P =10^{-3}k_BT/d^3$ and $P=0$. Under the isobaric process, the volume of a particle expands linearly with temperature but this manner is invalid at very low temperatures. At $T = 40$ and 60 $K$, $d \approx 0.177$ and 0.202 nm, respectively. The values are in the same order of size of the rhombus-like molecule in our simulations. 

Two approaches provide results relatively close to each other except for very high compression ($P = 2000$ bar). There are several possible reasons for this deviation: (1) roles of asymmetric structures of molecules on the glass transition become significant, meanwhile the ECNLE theory considers that the molecular shape is averagely spherical; (2) the effective range  of pressures within which $\lambda$ and $T_0$ could be treated as constant for the quasi-real system might be different than for real materials, i.e., the increase in pressure about 2000 bar could significantly affect on the correlation between local and collective dynamics for RLMS. The latter may be a main reason. However, to simplify theoretical calculations and minimize the number of parameters we ignore the pressure dependence of the correlation parameter $\lambda$. Good agreements between theory and simulation at a low pressure regime clearly validate this assumption. Quantitative comparisons of ECNLE theory with experiments at elevated pressures in prior works \cite{61,35} also support it.

Based on results presented in Fig. \ref{fig:1}, we can determine the glass transition temperature. The glass transition occurs when the time of experimental observation becomes comparable with the relaxation time of the system. This vitrification time is commonly equal to 100 seconds for the real materials \cite{67}. However, the time scales accessible in simulations are much shorter, and they are in ranges of nanoseconds. Thus, we define the glass transition temperatures as $\tau_\alpha(T_g)=1$ $ns$. Then, we can estimate pressure dependence of $T_g$, which is commonly parameterized for the isobaric dynamic fragility

\begin{eqnarray}
m = \left. \frac{\partial\log_{10}(\tau_\alpha)}{\partial(T_g/T)}\right |_{p,T=T_g}.
\label{eq:fragility}
\end{eqnarray}
 
\begin{figure}[htp]
\center
\includegraphics[width=9cm]{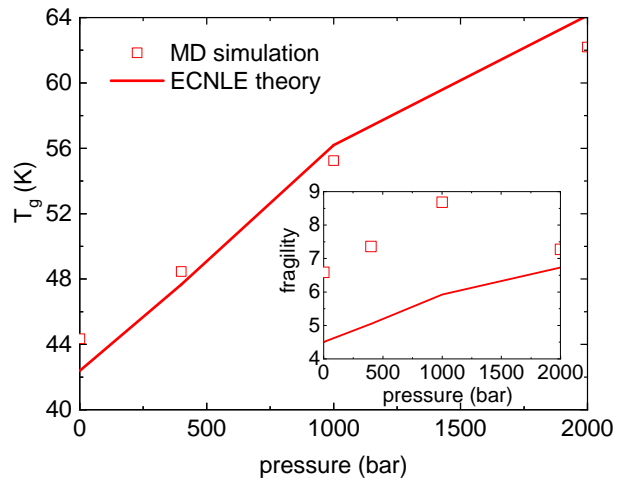}
\caption{\label{fig:3}(Color online) The glass transition temperature predicted by simulation and ECNLE theory as a function of pressure. The inset shows pressure effects on the dynamic fragility.}
\end{figure}

In Fig. \ref{fig:3}, one can see that the slowing down of molecular dynamics due to compression leads to an increase of $T_g$. Over a wide range of pressure, theoretical $T_g$s are close to simulation. Moreover, ENCLE theory predicts a monotonic growth of the dynamic fragility with compression, while simulation results suggest that this variation is non-monotonic and the fragility only increases when $P \leq 1000$ bar (see inset of Fig. \ref{fig:3}). The external pressure raises both local and collective elastic barriers in the ECNLE theory. The higher fragility implies that effects of collective motions on glassy dynamics is more important than those of the local dynamics. As a consequence, further investigation on the ECNLE theory would provide crucial information on the general behavior of the fragility. The mentioned above-reasons can explain why our statistical mechanical theory and simulations provide relatively different results. 

Although generally fragility decreases as pressure increases for van der Waals liquids, a variety of results are reported for other systems. The ECNLE prediction is consistent with several prior simulations \cite{14,15} and experiment \cite{24}. The behavior of our simulations agrees with effects of pressure on the structural relaxation in glycerol and xylitol \cite{25}. Meanwhile, the other widely exploited concept of the behavior of supercooled liquids dynamics, which is the density scaling law, predicts that isobaric fragility decreases during compression \cite{68}. Overall, this is an open and complicated issue due to many controversial conclusions.

\begin{figure}[htp]
\center
\includegraphics[width=9cm]{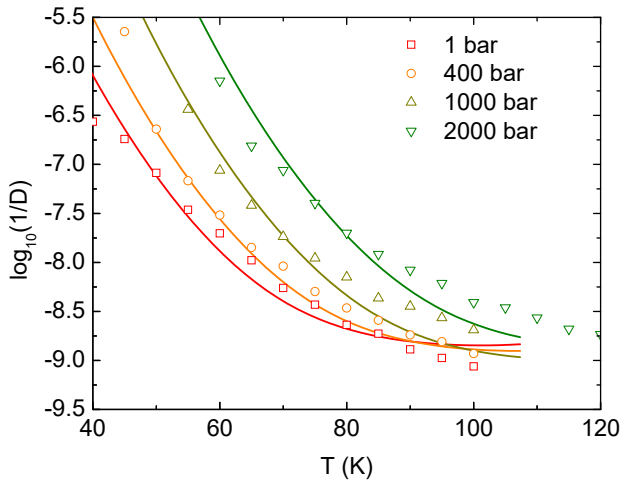}
\caption{\label{fig:2}(Color online) Logarithm of inverse diffusion constant (in unit of \ce{s/nm^2}) as a function of temperature at different pressures. Open points are simulations and solid curves correspond to ECNLE calculations.}
\end{figure}

Subsequently, by using the Eq. (\ref{eq:1}), we can estimate the temperature dependence of the inverse diffusion coefficients at various external pressures, which is shown in Fig. \ref{fig:2}. Again, ECNLE results agree well with simulations at pressures below 1000 bar. Additionally, one can observe intriguing behavior at higher temperatures, where predictions of ENCLE theory at different isobaric conditions intersect each other. Besides the previously mentioned reasons for inconsistency between ECNLE predictions and obtained simulation results for $\tau_\alpha$, there is another reason for the deviation. The ECNLE theory does not consider rotational motions, which are taken into account for calculations of $D$ from simulations results. If rotations are considered, the $\tau_\alpha$ is smaller, because of presents of additional degrees of freedom of molecules. It is consistent with consequences of  the incoherent intermediate scattering function definition, i.e., the rotational motions are essential for behavior of the incoherent intermediate scattering function for atoms, which decays faster then one for centers of molecules. Thus, the ECNLE theory overestimates $\tau_\alpha$ and consequently underpredicts $D$. 

Note that in Fig. \ref{fig:1} we show $\tau_\alpha$ calculated from simulation results for the center of molecules to directly verify the ECNLE predictions. However, in the case of real materials, as well as of studied herein quasi-real system, rotational motions influence on structural relaxation. One can expect that this effect is more evident at smaller densities where molecules can 'freely' rotate, i.e., at higher temperatures and smaller pressures. As a consequence, in Fig. \ref{fig:2}, predictions provided by the ECNLE theory deviate from simulations at higher temperatures, which are not observed for $\tau_\alpha$ in Fig. \ref{fig:1}.

\section{Relaxation-diffusion coupling}
Equation (\ref{eq:1}) connects $\tau_\alpha$ to $D$ with assumption that these two quantities are coupled. Based on this equation, we have
\begin{eqnarray}
\log_{10}\left(\frac{1}{D}\right) = \log_{10}6 + \log_{10}\tau_\alpha - 2\log_{10}(\lambda\Delta r).
\label{eq:9}
\end{eqnarray}

\begin{figure}[htp]
\center
\includegraphics[width=9cm]{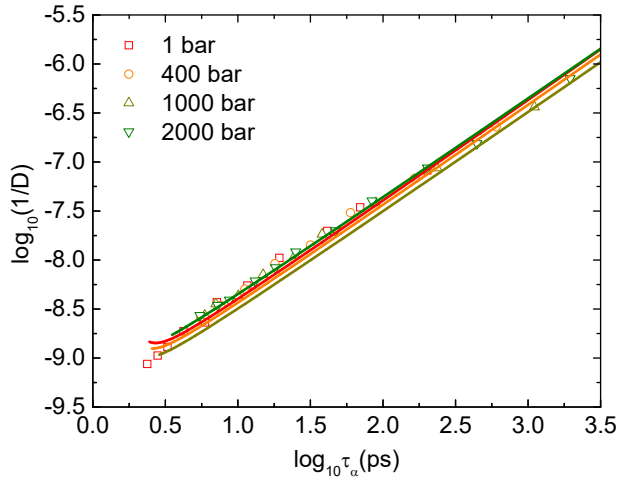}
\caption{\label{fig:4}(Color online) Logarithm of inverse diffusion constant (in unit of \ce{s/nm^2}) versus $\log_{10}\tau_\alpha$ at various pressures. Open points are simulations and solid curves correspond to ECNLE calculations.}
\end{figure}

Since $\Delta r \approx 0.2-0.4d$ in ECNLE calculations for various real materials \cite{2,7,8,10,6,42,11,61,62}, one obtains $\log_{10}(\lambda\Delta r) \ll \log_{10}\tau_\alpha$ and the linearity between $\log_{10}(1/D)$ and $\log_{10}\tau_\alpha$. To confirm this correlation, we use theoretical and simulation data in Fig. \ref{fig:1} and \ref{fig:2}, and contrast them in Fig. \ref{fig:4}. Remarkably, the double-log representation shows the almost perfect straight lines, which overlap each other. Their slopes are equal to 1.0 and weakly dependent on pressures. It is difficult to directly observe this behavior since the diffusion constant and alpha relaxation time can not be simultaneously measured in the same manner as theory and simulation.
This analysis can be deduced using behaviors of  typical liquids where translational motions of molecules are described by the standard diffusion equation. According to Maxwell's viscoelastic model, one finds that $\tau_\alpha$ is proportional to $\eta$ ($\eta$ is a viscosity). Meanwhile, the Stoke-Einstein equation reveals $D\propto1/\eta$ and thus $D\propto1/\tau_\alpha$.

Another analysis is based on the incoherent intermediate scattering function of a glass forming liquid, which is 
\begin{eqnarray}
F_S(q,t)=\frac{1}{N}\sum_{j=1}^N\left<e^{i\bf{q}(\bf{r}(0)-\bf{r}(t))} \right>
\label{eq:10}
\end{eqnarray}
where $N$ is the number of molecules. In the case of diffusion-relaxation coupling, $F_S\equiv\exp(-t/\tau_\alpha)$ can be expressed as $\exp(-q_0^2Dt)$, where $q_0$ is the wave vector corresponding to the position of the first peak in the static structure factor  \cite{69}. However, in the supercooled regime, $F_S(q_0,t)$ decays in the stretched-exponential way, which is described by $F_S(q_0,t)\sim\exp(-(t/\tau_\alpha)^{\beta_s})$. After obtaining $D$ and $\tau_\alpha$, we can determine the degree of the coupling by calculating their product because $D \tau_\alpha$ is constant when $D\propto 1/\tau_\alpha$.

\begin{figure}[htp]
\center
\includegraphics[width=9cm]{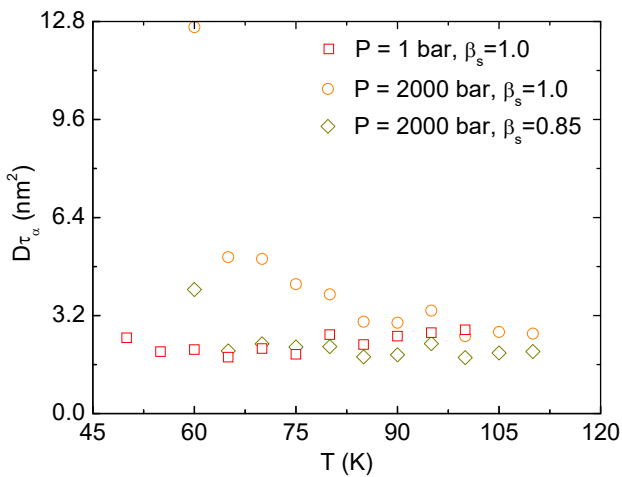}
\caption{\label{fig:5}(Color online) The temperature dependence of $D \tau_\alpha$ at $P = 1$ and 2000 bar obtained by simulation.}
\end{figure}

As it is shown in Fig. \ref{fig:5} at pressure equal to 1 bar, $D \tau_\alpha$ is constant over a whole range of studied temperatures. On the other hand, a significant increase of $D \tau_\alpha$ with cooling is observed at $P=2000$ bar. However, when considering the stretched-exponential decay of relaxation function with $\beta_s=0.85$, $D \tau_\alpha$ remains nearly unchanged for almost all temperature range. It implies that at high pressures the decoupling between $D$ and $\tau_\alpha$ could be expected, which has an additional influence on the ECNLE prediction for the diffusion constant.

\section{Conclusions}
We have shown simulations for RLMS to calculate the temperature and pressure dependence of alpha relaxation time and diffusion constant. Quantitative comparisons with the ECNLE theory are performed and numerical results demonstrate a good agreement, especially for relaxation times calculated for centers of the molecules. The observed deviations between simulation and theory at high pressures is due to anisotropy of the simulated molecular shape. The ECNLE theory employs a hard-sphere fluid to describe the glassy dynamics and influences of inter-molecular interactions and real molecular shape are attributed to the thermal mapping. Thus, precise agreement should not be expected. Another reason could be ignorance of rotational effects on the glass transition in the ECNLE theory. When pressure increases from 1 to 2000 bar, our theoretical calculations predict a monotonic growth of the dynamic fragility and glass transition temperature. These predictions are quantitatively consistent with simulations except for a decrease of the dynamic fragility at 2000 bar. Interestingly, all variations calculated by theory and simulation can be found in prior works \cite{14,15,24,25}. Based on these results, we find a coupling between the diffusion constant and structural relaxation time, which is $D \sim 1/\tau_\alpha$. The coupling manner remains unchanged in the compression process.



\begin{acknowledgments}
This work was supported by JSPS KAKENHI Grant Numbers JP19F18322 and JP18H01154. This research was funded by the Vietnam National Foundation for Science and Technology Development (NAFOSTED) under grant number 103.01-2019.318. KK and MP are deeply grateful for the financial support by the National Science Centre within the framework of the Maestro10 project (Grant No. UMO-2018/30/A/ST3/00323).
\end{acknowledgments}

\end{document}